\documentclass[aps,twocolumn,superscriptaddress,showpacs,showkeys,amsmath,amssymb]{revtex4}

\usepackage{amsfonts}
\usepackage{amssymb,amsmath}
\usepackage{graphicx}
\usepackage{dcolumn}
\usepackage{bm}
\usepackage{color}

\begin{document}

\title{Quantum phase transitions: The mean-field perspective}

\author{Johannes Richter}
\affiliation{Institut f\"{u}r theoretische Physik,
          Otto-von-Guericke-Universit\"{a}t Magdeburg,
          P.O. Box 4120, 39016 Magdeburg, Germany}

\author{Oleg Derzhko}
\affiliation{Institute for Condensed Matter Physics,
          National Academy of Sciences of Ukraine,
          Svientsitskii Street 1, 79011 L'viv, Ukraine}
\affiliation{Institut f\"{u}r theoretische Physik,
          Otto-von-Guericke-Universit\"{a}t Magdeburg,
          P.O. Box 4120, 39016 Magdeburg, Germany}
\affiliation{Department for Theoretical Physics,
          Ivan Franko National University of L'viv,
          Drahomanov Street 12, 79005 L'viv, Ukraine}
\affiliation{Abdus Salam International Centre for Theoretical Physics,
          Strada Costiera 11, 34151 Trieste, Italy}

\date{\today}



\begin{abstract}
To illustrate a simple mean-field-like approach for examining quantum phase transitions
we consider the $J-J^\prime$ quantum Heisenberg antiferromagnet on a square lattice.
The exchange couplings $J$ and $J^\prime$ are competing with each other. 
The ratio $J^\prime/J$ is the control parameter and its change drives the transition.
We adopt a variational ansatz, calculate the ground-state energy as well as the order parameter 
and describe the quantum phase transition inherent in the model.
This description corresponds completely to the standard Landau theory of phase transitions.
We also discuss how to generalize such an approach for more complicated quantum spin models.
\end{abstract}

\maketitle

\section{Thermal and quantum phase transitions}
\label{sec1}
\setcounter{equation}{0}

Phase transitions are ubiquitous.
Melting of solids, evaporation of liquids, disappearance of ferromagnetism upon heating 
are typical examples
to name just a few.

Thermodynamics and statistical physics provide a background for understanding phase transitions \cite{t_sp1,t_sp2,t_sp3}.
An important concept here is the order parameter.
Its behavior as varying some control parameter signalizes a phase transition.
For the ferromagnetic--paramagnetic phase transition driven by temperature (control parameter)
it is naturally to choose the total magnetization as the order parameter.
The magnetization is nonzero in the low-temperature ferromagnetic phase
but is zero in the high-temperature paramagnetic phase.
If the order parameter vanishes (or arises) continuously with varying of the control parameter
we face a continuous phase transition.

Statistical mechanics gives many exactly solvable microscopic models which exhibit phase transitions.
The square-lattice Ising model first solved by Lars Onsager in 1944 \cite{onsager}
is probably the most famous one.
Within the statistical mechanics picture,
the magnetization starts to fluctuate as the temperature deviates from zero 
resulting in the reduction of the ground-state magnetization.
As the temperature approaches the critical value,
the fluctuations are extremely developed
and the magnetization vanishes. 
Finally, it is zero for all temperatures above the critical temperature.

Interestingly,
there is a similar picture for a quantum many-particle system being in the ground state 
(i.e., at zero temperature)
where fluctuations have quantum nature. 
The temperature cannot serve as the control parameter. 
Rather, external pressure, magnetic field or competing terms in the Hamiltonian etc. 
may be appropriate to tune the strength of quantum fluctuations, 
i.e., these parameters can drive the transition at zero temperature.
The simplest example showing a quantum phase transition 
is the spin-1/2 Ising ferromagnet in a transverse magnetic field \cite{sachdev,voita,derzhko}
\begin{equation}
\label{101}
H= J\sum_{\langle nm\rangle} {s}^z_{n}{s}^z_{m}-h\sum_{n} s^x_{n},
\;\;\;
J<0,
\end{equation}
where the first sum runs over all nearest-neighbor pairs 
and the second sum over all lattice sites, $n=1,\ldots,N$.
While in the pure Ising model ($h=0$) no quantum fluctuations are present, 
the term with the transverse field does neither commute with the Ising interaction
nor with the operator of the order parameter $S^z=\sum_{n} {s}^z_{n}$, 
thus
introducing quantum fluctuations.  
Clearly, 
at zero field the ground state is the fully polarized ferromagnetic state 
with order parameter $\langle {\rm{GS}}\vert S^z\vert{\rm{GS}}\rangle=N/2$. 
Increasing  the field strength the magnetization (order parameter) first remains finite (although is reduced).
If the field strength approaches a critical value,
the fluctuations of the order parameter become extremely developed resulting in vanishing of the magnetization. 
Finally,   
above this critical field the order parameter is zero.
This picture has been confirmed experimentally \cite{bitko}.

Except the Ising model in the transverse field 
there are many other quantum spin models exhibiting quantum phase transitions.
In particular, 
spin-1/2 Heisenberg models with competing bonds provide a large variety of models 
which are often appropriate to describe experimental findings in magnetic compounds \cite{buch1,buch2}.
Below we will discuss  such a quantum spin Heisenberg model in two dimensions in some detail.   
For this purpose we use the variational approach
which is a widely used tool in theoretical physics in general \cite{ritz}
and in the theory of quantum many-body systems in particular \cite{c_gros,becca}.

The rest of the paper is organized as follows.
First we describe the model used to study a quantum phase transition,
Sec.~\ref{sec2}.
Next, we introduce a variational ansatz and determine observables, 
Sec.~\ref{sec3A}.
The elaborated theory can be cast into the standard Landau theory of phase transitions, 
Sec.~\ref{sec3B}. 
Then we discuss some generalizations, 
Sec.~\ref{sec4}.
Finally, we summarize our findings and sketch perspectives for further work, 
Sec.~\ref{sec5}.

\section{The $J-J^\prime$ quantum Heisenberg antiferromagnet}
\label{sec2}
\setcounter{equation}{0}

\begin{figure}
\begin{center}
\includegraphics[clip=on,width=60mm,angle=0]{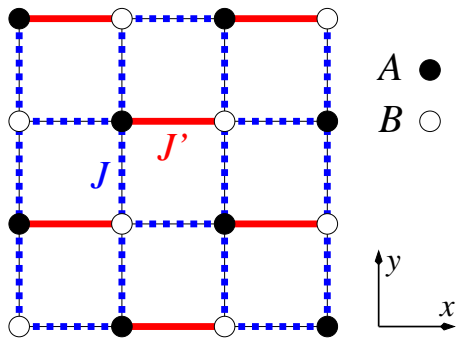}
\caption
{(Color online) 
Square-lattice $J-J^\prime$ model.
The square lattice consists of two sublattices to be denoted as $A$ and $B$.
$J^\prime$ bonds form a staggered dimer pattern.}
\label{fig01}
\end{center}
\end{figure}

As already mentioned in the Introduction 
a canonical model to study quantum phase transitions 
is the spin-1/2 Heisenberg model with competing exchange bonds, say $J$ and $J^\prime$,  
\begin{eqnarray}
\label{201}
H
=
J\sum_{\langle nm\rangle} {\bf s}_{n}\cdot {\bf s}_{m}
+
J^\prime\sum_{\langle nm\rangle^\prime} {\bf s}_{n}\cdot {\bf s}_{m} .
\end{eqnarray}
Here the control parameter typically is the ratio of $J$ and $J^\prime$.
In what follows we call model (\ref{201}) the $J-J^\prime$ model. 
Let us consider a specific example for such a $J-J^\prime$ model
namely  a square-lattice model 
with two different antiferromagnetic nearest-neighbor interactions $J>0$ and $J^\prime>0$
as shown in Fig.~\ref{fig01}.
Both $J$ and $J^\prime$ are positive, 
we also assume that $J^\prime \ge J$,
and the $J^\prime$ bonds form a staggered  
(in contrast to, say, columnar or herringbone, see Refs.~\onlinecite{wenzel,fritz}) 
covering of the square lattice.

\begin{figure}
\begin{center}
\includegraphics[clip=on,width=80mm,angle=0]{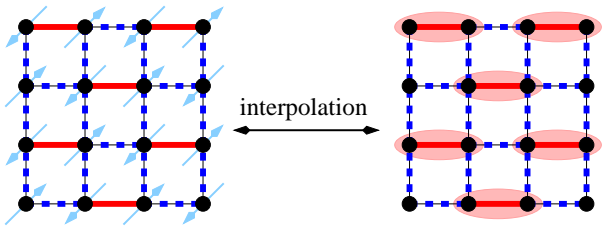}
\caption
{(Color online) 
Left: Pictorial representation of the N\'{e}el state.
Right: Pictorial representation of the singlet-product state.}
\label{fig02}
\end{center}
\end{figure}

If $J^\prime=J$ we face the well-investigated square-lattice spin-half Heisenberg antiferromagnet, 
see, e.g., Refs.~\onlinecite{manousakis,oitmaa1992,darradi2008}.
In this limit the ground state exhibits antiferromagnetic N\'{e}el-type long-range order.
It is important to notice that the simple N\'{e}el product state 
$\Psi_{\mbox{N\'{e}el}}=\vert\uparrow\rangle\vert\downarrow\rangle\vert\uparrow\rangle\vert\downarrow\rangle\ldots$ 
(as shown pictorially on the left side of Fig.~\ref{fig02})
is not an eigenstate of the quantum model. 
Rather, the ground state is a more complex many-body state with N\'{e}el-type long-range order, 
where the sublattice magnetization is reduced by quantum fluctuations to about 60\% of the classical value \cite{manousakis,oitmaa1992,darradi2008}.
Since the main features of the classical N\'{e}el order are present in the quantum  model as well, 
we call this type of order semiclassical N\'{e}el order.

On the other hand, 
in the limit of $J^\prime/J \to \infty$ the ground state is of quantum nature without a classical reference state.
This can be easily seen by considering an isolated spin pair coupled by the antiferromagnetic bond $J^\prime$: 
While for the Ising model (and similarly for the classical Heisenberg model) 
a single state $\vert\uparrow\rangle\vert\downarrow\rangle$ 
(or equivalently $\vert\downarrow\rangle\vert\uparrow\rangle$)
can serve as the ground state, 
for the quantum spin-$1/2$ Heisenberg case
only the superposition of both to a spin singlet, 
i.e., 
$(\vert\uparrow\rangle\vert\downarrow\rangle - \vert\downarrow\rangle\vert\uparrow\rangle)/\sqrt{2}$, 
is the ground  state.
For the lattice, at $J^\prime/J \to \infty$ the ground state is the regular pattern of singlets 
(on the bonds of strength $J^\prime$, 
as shown pictorially on the right side of Fig.~\ref{fig02}), 
which we call valence-bond state. 
That is a pure quantum state with zero sublattice magnetization.

Thus, we expect that at some critical value of the ratio $J^\prime/J>1$,
a phase transition between the magnetically ordered N\'{e}el phase and the valence-bond singlet phase without magnetic order occurs.
Let us emphasize that for the Ising model and also for classical Heisenberg spins
(i.e., ${\bf{s}}_n$ stands for the classical vector of the length $s$) 
the  N\'{e}el state is the ground state for all values of antiferromagnetic $J^\prime$ and $J$, 
and no transition takes place at all.
In what follows we take the spin-$1/2$  $J-J^\prime$ model on the square lattice 
as a paradigm to elaborate a mean-field-like description of quantum phase transitions \cite{lnp,gros,sven,johannes,rashid1,rashid2}.

\section{The variational approach: Interpolating between the N\'{e}el state and the valence-bond state}
\label{sec3}
\setcounter{equation}{0}

\subsection{Variational ansatz, the ground-state energy, and the order parameter}
\label{sec3A}

To interpolate between the two limiting cases,
$J^\prime/J=1$ and $J^\prime/J\gg 1$,
we introduce the following variational state \cite{gros,lnp,sven,johannes,rashid1,rashid2}
\begin{eqnarray}
\label{301}
\vert\Psi_{\rm{var}}\rangle
=
\prod_{i\in A}\frac{1}{\sqrt{1+t^2}}
\left(
\vert \uparrow_i \downarrow_{i+\hat{x}} \rangle - t \vert \downarrow_i \uparrow_{i+\hat{x}} \rangle
\right).
\end{eqnarray}
The two lattice sites $i$ and $i+\hat{x}$ in the r.h.s. of Eq.~(\ref{301}) correspond to a $J^\prime$ bond
and thus the product is taken over all $J^\prime$ bonds of the lattice, see Fig.~\ref{fig01}.

Furthermore, 
$0\le t\le 1$
is the variational parameter
to be determined from the minimum condition for the variational energy
$E(t)=\langle \Psi_{\rm{var}} \vert H\vert \Psi_{\rm{var}} \rangle$.
If $t=0$ Eq.~(\ref{301}) gives the N\'{e}el state;
if $t=1$ Eq.~(\ref{301}) gives the valence-bond singlet state 
(i.e., the product state of the singlets on $J^\prime$ bonds).

\begin{figure}
\begin{center}
\includegraphics[clip=on,width=60mm,angle=0]{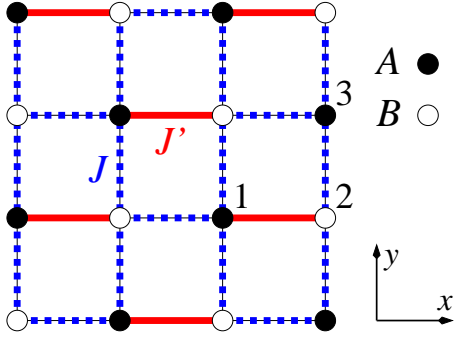}
\caption
{(Color online) 
Numeration of sites, see also Fig.~\ref{fig01}, introduced for computation purposes.}
\label{fig03}
\end{center}
\end{figure}

Let us calculate the variational energy $E(t)=\langle \Psi_{\rm{var}} \vert H\vert \Psi_{\rm{var}} \rangle$.
We choose the numeration of sites as shown in Fig.~\ref{fig03},
i.e., 
the sites 1 and 2 are connected by a $J^\prime$ bond
and 
the sites 2 and 3 are connected by a $J$ bond.
We get
\begin{eqnarray}
\label{302}
&&\left(s^+_1s^-_2 + s^-_1s^+_2\right)\vert \Psi_{\rm{var}} \rangle =
\nonumber\\
&&\qquad \ldots
\frac{1}{\sqrt{1+t^2}}\left(\vert\downarrow_1\uparrow_2\rangle - t\vert\uparrow_1\downarrow_2\rangle\right)
\ldots,
\nonumber\\
&&s^z_1s^z_2\vert \Psi_{\rm{var}} \rangle =
\nonumber\\
&&\qquad \ldots
\left(-\frac{1}{4}\right)
\frac{1}{\sqrt{1+t^2}}
\left(\vert\uparrow_1\downarrow_2\rangle - t\vert\downarrow_1\uparrow_2\rangle\right)
\ldots,
\nonumber\\
&&\langle \Psi_{\rm{var}} \vert {\bf{s}}_1\cdot{\bf{s}}_2 \vert \Psi_{\rm{var}} \rangle
=
-\frac{t}{1+t^2}-\frac{1}{4}
\end{eqnarray}
and
\begin{eqnarray}
\label{303}
&&\left(s^+_2s^-_3 + s^-_2s^+_3\right)\vert \Psi_{\rm{var}} \rangle =
\nonumber\\
&&\quad \ldots
\frac{1}{\sqrt{1+t^2}}\vert\uparrow_1\uparrow_2\rangle 
\frac{1}{\sqrt{1+t^2}}\vert\downarrow_3\downarrow_{3+\hat{x}}\rangle
\ldots +
\nonumber\\
&& \quad
\ldots
t^2
\frac{1}{\sqrt{1+t^2}}\vert\downarrow_1\downarrow_2\rangle 
\frac{1}{\sqrt{1+t^2}}\vert\uparrow_3\uparrow_{3+\hat{x}}\rangle
\ldots,
\nonumber\\
&&s^z_2s^z_3\vert \Psi_{\rm{var}} \rangle =
\nonumber\\
&&\quad
\ldots
\left(-\frac{1}{4}\right)
\frac{1}{\sqrt{1+t^2}}\left(\vert\uparrow_1\downarrow_2\rangle + t\vert\downarrow_1\uparrow_2\rangle\right)
\times
\nonumber\\
&&\quad 
\frac{1}{\sqrt{1+t^2}}\left(\vert\uparrow_3\downarrow_{3+\hat{x}}\rangle + t\vert\downarrow_3\uparrow_{3+\hat{x}}\rangle\right)
\ldots,
\nonumber\\
&&\langle \Psi_{\rm{var}} \vert {\bf{s}}_2\cdot{\bf{s}}_3 \vert \Psi_{\rm{var}} \rangle
=
-\frac{1}{4}\left(\frac{1-t^2}{1+t^2}\right)^2.
\end{eqnarray}
Combining Eqs.~(\ref{302}) and (\ref{303}) we find
\begin{eqnarray}
\label{304}
\frac{E(t)}{N}&=&\frac{\langle \Psi_{\rm{var}} \vert H \vert \Psi_{\rm{var}} \rangle}{N}
=
\nonumber\\
&&
-\frac{J^\prime}{2}\frac{t}{1+t^2}
-\frac{J^\prime}{8}
-\frac{(z-1)J}{8}\left(\frac{1-t^2}{1+t^2}\right)^2,
\quad
\end{eqnarray}
where $z$ is the number of nearest neighbors, 
i.e., $z=4$ for the square lattice.

Next task is to find the value of $t$ which yields the minimum of $E(t)$.
Since
\begin{eqnarray}\
\label{305}
\frac{{\rm{d}}}{{\rm{d}}t}\frac{E(t)}{N}
=
\frac{J^\prime t^4 -6Jt^3 +6Jt -J^\prime}{2(1+t^2)^3},
\end{eqnarray}
we get a fourth order algebraic equation with respect to $t$:
\begin{eqnarray}
\label{306}
&&J^\prime t^4 -6Jt^3 +6Jt-J^\prime =
\nonumber\\
&& \quad \left(J^\prime t^2 -6Jt +J^\prime\right)\left(t^2-1\right)=0.
\end{eqnarray}
Equation~(\ref{306}) has the following solutions:
\begin{eqnarray}
\label{307}
t_1&=&-1,
\nonumber\\
t_2&=&\frac{3J}{J^\prime} -\sqrt{\left(\frac{3J}{J^\prime}\right)^2-1},
\nonumber\\
t_3&=&1,
\nonumber\\
t_4&=&\frac{3J}{J^\prime} +\sqrt{\left(\frac{3J}{J^\prime}\right)^2-1}.
\end{eqnarray}
The solutions $t_1$ and $t_3$ exist for all $J^\prime$,
whereas the solutions $t_2$ and $t_4$ are real for $J^\prime\le 3J$ only.
However, we have to discard the solutions $t_1$ and $t_4$, 
because they do not obey the imposed restriction $0\le t\le 1$.
Now, $E(t)/N$ is minimal for 
\begin{eqnarray}
\label{308}
t
=
\left\{
\begin{array}{ll}
\frac{3J}{J^\prime} -\sqrt{\left(\frac{3J}{J^\prime}\right)^2-1}, & J^\prime\le 3J, \\
1,                                                                & J^\prime > 3J.
\end{array}
\right.
\end{eqnarray}
Thus, the ground-state energy (per site) is given by
\begin{eqnarray}
\label{309}
\frac{E_0}{N}
=
\left\{
\begin{array}{ll}
-\frac{3J^\prime}{8}-\frac{3J}{8}\left(1-\frac{J^\prime}{3J}\right)^2, & J^\prime\le 3J, \\
-\frac{3J^\prime}{8},                                                  & J^\prime > 3J.
\end{array}
\right.
\end{eqnarray}

\begin{figure}
\begin{center}
\includegraphics[clip=on,width=80mm,angle=0]{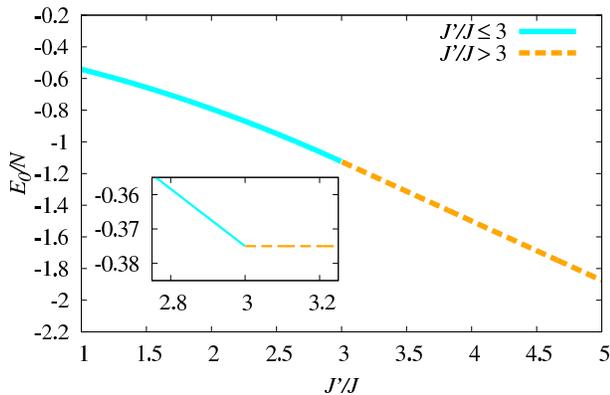}
\caption
{(Color online) 
The  $E_0/N$ versus $J^\prime/J$, see Eq.~(\ref{309}).
Inset: The first derivative of the ground-state energy per site, 
${\rm{d}}(E_0/N)/{\rm{d}}J^\prime$, 
in dependence on  $J^\prime/J$.}
\label{fig04}
\end{center}
\end{figure}

Following Ehrenfest's classification of phase transitions \cite{t_sp1},
we may inspect the derivatives of the ground-state energy 
(that plays here the role of the relevant thermodynamic potential) 
with respect to the control parameter $J^\prime$. 
Easily we find that ${\rm{d}}E_0/{\rm{d}}J^\prime$ is continuous everywhere, 
but the second derivative, ${\rm{d}}^2E_0/{\rm{d}}{J^\prime}^2$, has a jump at $J^\prime=3J$,
see Fig.~\ref{fig04}.
Therefore, we have a first indication that there is a continuous quantum phase transition driven by $J^\prime/J$.

To confirm this finding,
we calculate the variational sublattice magnetization (per site)
$m(t)=\langle\Psi_{\rm{var}}\vert s_1^z\vert \Psi_{\rm{var}}\rangle$,
that will yield the relevant order parameter.
Since
\begin{eqnarray}
\label{310} 
s_1^z\vert \Psi_{\rm{var}}\rangle
=
\ldots
\frac{1}{2}\frac{1}{\sqrt{1+t^2}}
\left(\vert\uparrow_1\downarrow_2\rangle + t \vert\downarrow_1\uparrow_2\rangle\right)
\ldots,
\end{eqnarray}
we arrive at
\begin{eqnarray}
\label{311}
m(t)=\frac{1}{2}\frac{1-t^2}{1+t^2}
\end{eqnarray}
for the variational magnetization at a site on the sublattice $A$.
For the corresponding magnetization at a site on the sublattice $B$ the same expression, but with the opposite sign, is valid.
Inserting the optimal $t$ from Eq.~(\ref{308}) we get for the order parameter
\begin{eqnarray}
\label{312}
m_0
=
\left\{
\begin{array}{ll}
\frac{1}{2}
\sqrt{\left(1+\frac{J^\prime}{3J}\right)\left(1-\frac{J^\prime}{3J}\right)}, & J^\prime\le 3J, \\
0,                                                                           & J^\prime > 3J.
\end{array}
\right.
\end{eqnarray}
In accordance with the findings for the ground-state energy, 
Eq.~(\ref{312}) yields a continuous transition with a quantum critical point at $J^\prime_c=3J$ 
and a critical exponent $\beta=1/2$ obvious from the behavior of $m_0$ as $J^\prime \to J^\prime_c-0$. 
Moreover, 
we have $m_0=\sqrt{2}/3\approx 0.471$ at $J^\prime =J$, 
i.e., already within our simple approach the sublattice magnetization is reduced by quantum fluctuations compared to its classical value $m_0^{{\rm{class}}}=1/2$.
Knowing the variational wave function (\ref{301}) with the variational parameter given in Eq.~(\ref{308})
we are able to calculate any observable quantity.

As reported above the critical index of the order parameter is that of a mean-field theory.
The question arises how the mean-field character of our approach is evident.
The crucial point is the product form of our wave function (\ref{301}),
i.e.,
there is no mutual correlation between the individual species of the system
(i.e., the dimers on the $J^\prime$ bonds) 
in our wave function.

\subsection{Landau theory}
\label{sec3B}

Due to the mean-field character of our approach it is natural to ask whether
the famous Landau theory \cite{landau1937,t_sp1} 
is applicable to describe the critical behavior discussed above.     
Indeed,
our approach can be cast into the standard Landau theory of phase transitions.
The starting point of the Landau theory is the expansion of the (variational) free energy as a function of the order parameter.
Then the free energy should be minimized with respect to the order parameter.
In our case,
we need an expansion of the ground-state energy in powers of the sublattice magnetization. 
To get such an expansion,
we use Eq.~(\ref{311}) to express $t$ in terms of $m$,
that is,
$t(m)=\sqrt{(1-2m)/(1+2m)}$.
We substitute $t(m)$ into Eq.~(\ref{304}) and get
\begin{eqnarray}
\label{313}
\frac{E(m)}{N}
=
-\frac{J^\prime}{4}\sqrt{1-4m^2}
-\frac{J^\prime}{8}
-\frac{3}{2}J m^2.
\end{eqnarray}
Then expanding $E(m)/N$ (\ref{313}) in powers of $m$ yields
\begin{eqnarray}
\label{314}
\frac{E(m)}{N}
=
-\frac{3}{8}J^\prime
+\frac{1}{2}\left(J^\prime - 3J\right)m^2
+\frac{1}{2}J^\prime m^4
+\ldots .
\end{eqnarray}
This is the variational ground-state energy (per site) as a function of the (small) order parameter $m$.

Within Landau's theory of thermal phase transitions \cite{landau1937,t_sp1},
the simplest 
(i.e., the case of a scalar order parameter $m$)
starting point is the following (variational) free energy expansion:
\begin{eqnarray}
\label{315}\
F(T,m)&=&F(T,m=0) + A(T)m^2 +Bm^4,
\nonumber\\
A(T)&=&a(T-T_c),
\;\;\;
a>0,
\;\;\;
B>0.
\end{eqnarray}
At the critical temperature $T=T_c$,
the coefficient $A(T)$ changes its sign
resulting in a qualitative change of the dependence $F(T,m)$.
The order parameter $m$ must realize the minimum of $F(T,m)$ and therefore it has the following temperature dependence:
\begin{eqnarray}
\label{316}
m_0(T)
\propto
\left\{
\begin{array}{ll}
\sqrt{T_c-T}, & T\le T_c, \\
\hspace{1.2cm} 0,            & T > T_c.
\end{array}
\right.
\end{eqnarray}
Thus, the critical exponent $\beta=1/2$.

\begin{figure}
\begin{center}
\includegraphics[clip=on,width=80mm,angle=0]{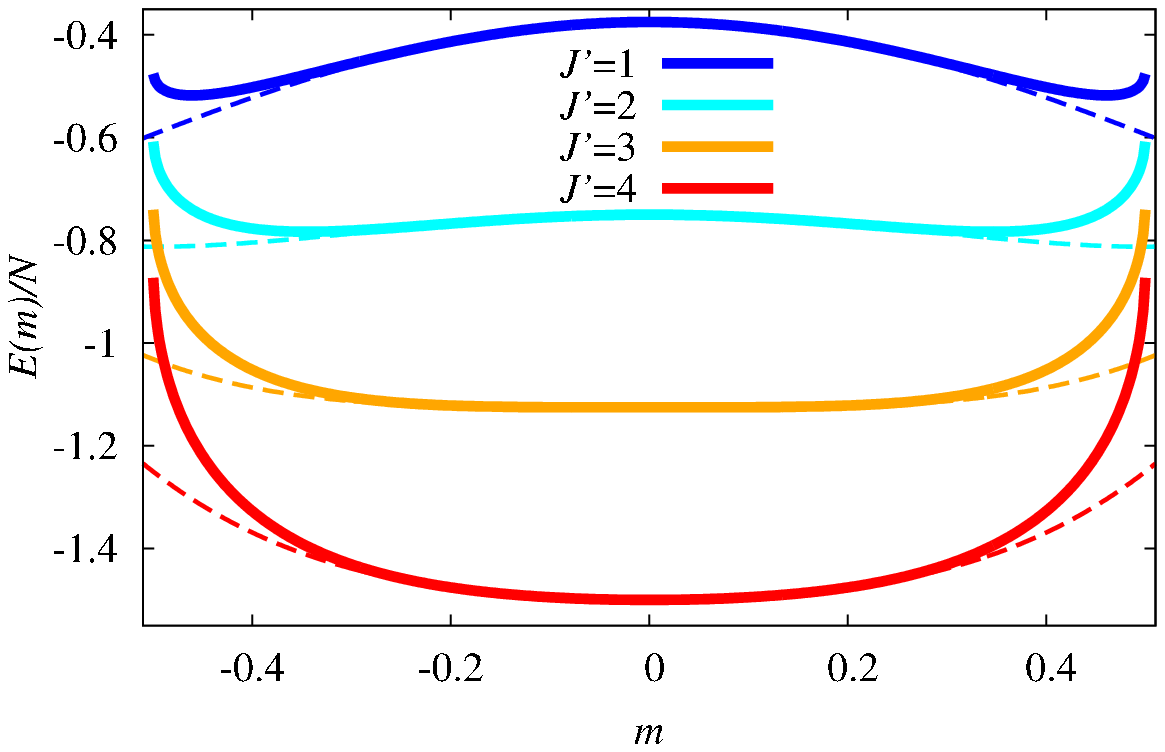}\\
\vspace{3mm}
\includegraphics[clip=on,width=80mm,angle=0]{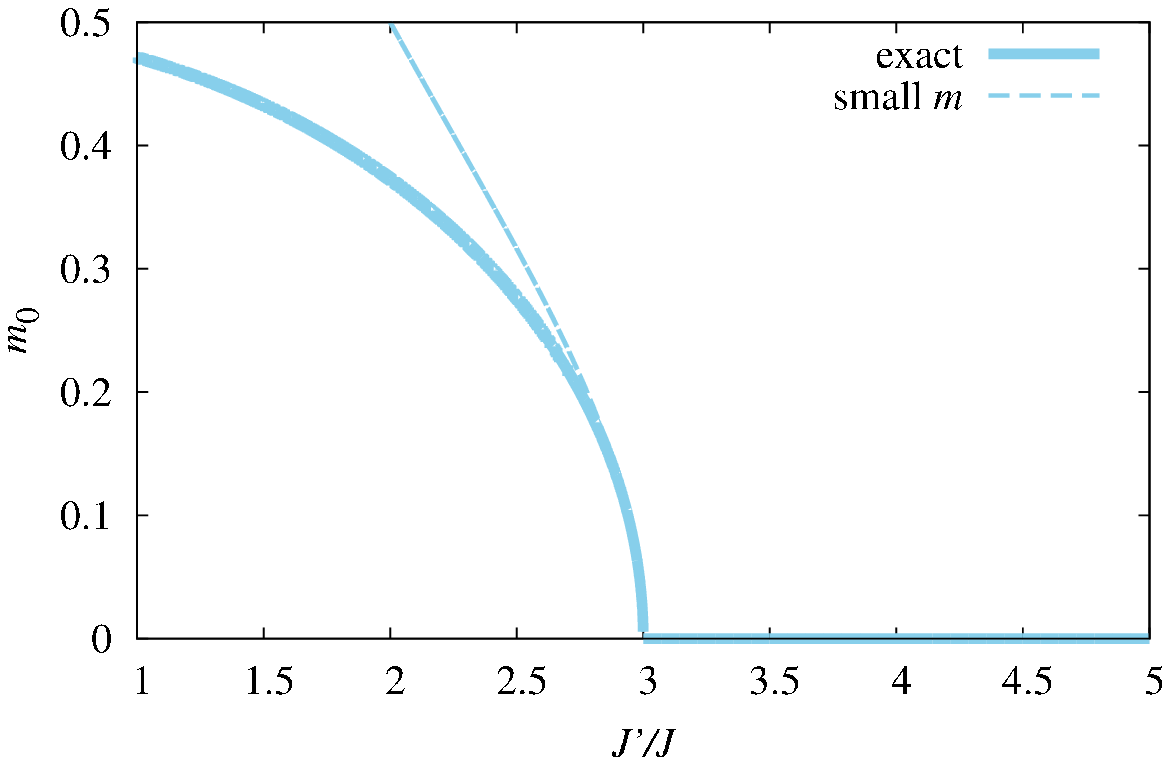}
\caption
{(Color online) 
Upper panel: 
The variational ground-state energy $E/N$ versus $m$ at different values of the control parameter $J^\prime/J$.
Thin dashed lines correspond to Eq.~(\ref{314}) (small $m$ approximation) 
and 
thick solid lines correspond to Eq.~(\ref{313}) (full expression for arbitrary $m$).
Lower panel:
The order parameter $m_0$ as a function of the control parameter $J^\prime/J$.
The thin dashed line corresponds to Eq.~(\ref{317}) (small $m$ approximation) 
and 
the thick solid line corresponds to Eq.~(\ref{312}) (full expression for arbitrary $m$).}
\label{fig05}
\end{center}
\end{figure}

Clearly,
Eq.~(\ref{314}) is a counterpart of Eq.~(\ref{315}).
In the upper panel of Fig.~\ref{fig05} 
we illustrate the qualitative change of the dependence $E(m)$,
cf. Eqs.~(\ref{313}), (\ref{314}),
as the control parameter $J^\prime/J$ crosses the critical value $J^\prime_c/J=3$.
Furthermore,
in the lower panel of Fig.~\ref{fig05} 
we illustrate the dependence of $m_0$ on $J^\prime/J$,
\begin{eqnarray}
\label{317}
m_0
=
\left\{
\begin{array}{ll}
\sqrt{\frac{3J-J^\prime}{2J^\prime}}, & J^\prime\le 3J, \\
\hspace{1.1cm} 0,                                    & J^\prime > 3J,
\end{array}
\right.
\end{eqnarray}
as the control parameter $J^\prime/J$ crosses the critical value $J^\prime_c/J=3$.
Note that Eq.~(\ref{317}) corresponds to Eq.~(\ref{316}), and, moreover, it
agrees with Eq.~(\ref{312}):
The latter equation transforms into the former one if we assume that $J^\prime\to 3J-0$.

Exploiting the relation of our approach to the Landau theory 
we can go one step forward
and consider now the effects of the spatial variation of the order parameter within Landau-Ginzburg theory.
The spatially dependent free-energy density now contains 
besides the local term attached to $J^\prime$-bonds corresponding to (\ref{315}) (however, with a space-dependent $m$)
also the non-local gradient term stemmed from $J$-bonds that is proportional to $\vert\nabla m\vert^2$.

\begin{figure}
\begin{center}
\includegraphics[clip=on,width=60mm,angle=0]{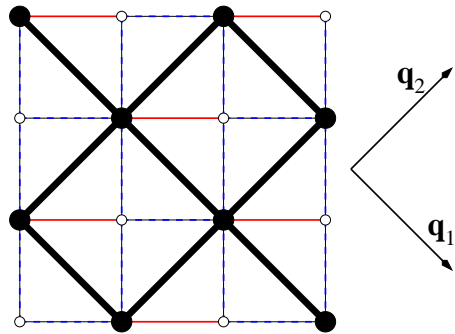}
\caption
{(Color online) 
Towards Eq.~(\ref{318}).
If the site 1 from Fig.~\ref{fig03} is denoted now by ${\bf{r}}$ 
then the site 3 from Fig.~\ref{fig03} is ${\bf{r}}+{\bf{q}}_2$.}
\label{fig06}
\end{center}
\end{figure}

Hence, we allow the variational parameter $t$ in Eq.~(\ref{301}) to be spatially dependent.
Equivalently,
we may assume the variational sublattice magnetization (per site) $m$ to be space dependent
since both quantities are tied together by Eq.~(\ref{311}).
After relaxing the condition of uniform $m$,
we have to reconsider Eq.~(\ref{313}) for the variational ground-state energy $E(m)$
which becomes now a functional of $m({\bf{r}})$.
Recalling its derivation
we conclude that now
\begin{eqnarray}
\label{318}
\frac{E[m({\bf{r}})]}{N}
=
-\frac{J^\prime}{4{\cal{N}}}\sum_{{\bf{r}}}\left(\sqrt{1-4m^2({\bf{r}})} + \frac{1}{2}\right)
-\frac{J}{2{\cal{N}}}
\times
\nonumber\\
\sum_{{\bf{r}}}
m({\bf{r}})
\left(m({\bf{r}}+{\bf{q}}_1)+m({\bf{r}}+{\bf{q}}_1+{\bf{q}}_2)+m({\bf{r}}+{\bf{q}}_2)\right).
\end{eqnarray}
Here ${\bf{r}}$ runs over ${\cal{N}}=N/2$ sites of the square lattice defined by, say, the left sites of the dimer bonds, see Fig.~\ref{fig06}.
It is convenient to rearrange the terms in the second sum in Eq.~(\ref{318}) replacing
$m({\bf{r}})m({\bf{r}}+{\bf{q}}_1)$
by 
$(m({\bf{r}}-{\bf{q}}_1)m({\bf{r}})+m({\bf{r}})m({\bf{r}}+{\bf{q}}_1))/2$
and so on.
We assume $m$ to be a slowly varying function.
Therefore we can expand
$m({\bf{r}}+{\bf{q}}_1)\approx m({\bf{r}})+{\bf{q}}_1\cdot \nabla m ({\bf{r}})+ {\bf{q}}_1\cdot \nabla ({\bf{q}}_1\cdot \nabla m ({\bf{r}}))/2$
and so on.
Furthermore,
we may replace the sum by the integral:
$(1/{\cal{N}})\sum_{{\bf{r}}}(\ldots) 
\to
(1/V)\int_V{\rm{d}}{\bf{r}}(\ldots)$
where $V=(\sqrt{2}a)^2{\cal{N}}$ and $a$ is the edge length of the square-lattice cell in Fig.~\ref{fig01}.
While in the first term in the r.h.s. in Eq.~(\ref{318})
we have simply to expand the square root in powers of $m$,
$\sqrt{1-4m^2}\to 1-2m^2-2m^4$,
in the second term we have to 
integrate by parts, 
neglect the boundary terms, 
insert ${\bf{q}}_1=\sqrt{2}a(1,0)$,  ${\bf{q}}_2=\sqrt{2}a(0,1)$,
and take into account that $(\nabla m({\bf{r}}))_x=(\nabla m({\bf{r}}))_y$.
Finally, we arrive at the following result:
\begin{eqnarray}
\label{319}
\frac{E[m({\bf{r}})]}{N}
=
\frac{1}{V}\int_V{\rm{d}}{\bf{r}}\, e(m({\bf{r}})),
\nonumber\\
e(m({\bf{r}}))
=-\frac{3J^\prime}{8}+\frac{J^\prime-3J}{2} m^2({\bf{r}})+\frac{J^\prime}{2} m^4({\bf{r}})
\nonumber\\
+\frac{3a^2J}{2}\vert \nabla m({\bf{r}}) \vert^2,
\end{eqnarray}
i.e., at a field theory.
It should be underlined that Eq.~(\ref{319}) does not represent the true field theory of the model (\ref{201}), 
see Sec.~\ref{sec3C},
since it is restricted only to the imposed variational states given in Eq.~(\ref{301}).

Landau-Ginzburg theory,
when the term with the fourth power is neglected,
allows to obtain the correlation function $\langle m({\bf{r}}_1) m({\bf{r}}_2)\rangle$.
It decays exponentially,
$\propto \exp(-\vert {\bf{r}}_1- {\bf{r}}_2 \vert/\xi)$,
with the correlation length $\xi$.
The correlation length diverges at the critical point $J^\prime_c=3J$ with the exponent $\nu=1/2$,
i.e., $\xi\propto 1/\sqrt{\vert J^\prime-J^\prime_c\vert}$.

\subsection{Beyond the mean-field treatment}
\label{sec3C}

The model discussed above at the mean-field level
has been studied using more sophisticated approaches.
Remarkably,
it has attracted some interest recently because of a suspicious of a new universality class \cite{wenzel,fritz}.
Let us explain this issue in more detail.

S.~Wenzel, L.~Bogacz, and W.~Janke studied the spin-1/2 $J-J^\prime$ square-lattice Heisenberg antiferromagnet 
with the staggered arrangement of $J^\prime$ bonds 
(i.e., the $J^\prime$-bond pattern as in Fig.~\ref{fig01})
by means of (stochastic series expansion) quantum Monte Carlo simulations \cite{wenzel}.
To probe the nature of the quantum phase transition, 
they calculated several well-known observables such as the staggered magnetization, the correlation length, and the spin stiffness.
All observables indicate a single phase transition 
and the critical point is determined as $J_c^\prime/J=2.5196(2)$.
Note that the critical point, 
$J_c^\prime/J=3$,
obtained by our simple variational ansatz (\ref{301}),  
is in reasonable agreement with this number.
However, 
in contrast to a general belief, 
they found that the critical exponent $\nu$ obtained by analyzing the data for all observables according to a finite-size scaling ansatz is $\nu=0.689(5)$.
This quantity is smaller 
than the standard three-dimensional classical Heisenberg model [i.e., $O(3)$] universality class
$\nu=0.7112(5)$.
This contradicts a common wisdom:
Quantum phase transition can be mapped onto classical phase transition in one higher dimension
and, in general, one expects 
that the quantum phase transition in $D$ space dimensions 
is in the universality class of the $(D+1)$-dimensional classical model \cite{sachdev,wenzel,fritz}.
In the second paper \cite{wenzel},
S.~Wenzel and W.~Janke extended their studies for more geometric arrangements of competing $J^\prime$ bonds
and confirmed that the critical exponent $\nu$ for other considered coupled-dimer magnets
is in excellent agreement with the classical $O(3)$ universality class.

A resolution to this puzzle put forward by the numerics of Ref.~\onlinecite{wenzel}
was proposed by L.~Fritz et al. \cite{fritz}.
They showed that there are indeed two different classes of coupled-dimer magnets.
While the first class
(consisting, in particular, of the columnar-dimer \cite{wenzel} or bilayer square-lattice systems)
follows the standard $O(3)$ universality class,
the low-energy quantum field theory of the other class
(consisting, in particular, of the staggered-dimer \cite{wenzel} or herringbone-dimer square-lattice systems)
is characterized by an additional cubic interaction of critical fluctuations, which has no classical analog.
As a result,
the asymptotic critical exponents are of the $O(3)$ universality class, 
but anomalously large corrections to scaling arise from this cubic interaction term.
The authors of Ref.~\onlinecite{fritz} also presented quantum Monte Carlo simulations 
that can be consistently interpreted in terms of critical exponents of the standard $O(3)$ universality class,
but with anomalously large corrections to scaling.

Clearly, 
the discussion of Refs.~\onlinecite{wenzel,fritz} 
which is based on such refined techniques as quantum Monte Carlo, effective low-energy quantum field theory and renormalization-group analysis etc.
is unreachable at the mean-field level.
The mean-field approach based on Eq.~(\ref{301}) cannot differ between the two different classes of coupled-dimer magnets.

\section{Generalizations}
\label{sec4}
\setcounter{equation}{0}

Our variational approach allows a straightforward discussion of modifications of the $J-J^\prime$ model.
In particular, 
we may modify the model in such a way that we can influence the strength of quantum fluctuations, 
e.g.,  
by anisotropy, 
by increasing the spin quantum number $s$, 
or 
by adding frustrating next-nearest-neighbor interactions.   
In what follows we will present a brief discussion of such modifications. 
 
\subsection{The easy-axis $XXZ$ $J-J^\prime$ model}

First we consider the anisotropic $XXZ$ model 
[instead of the isotropic $XXX$ Heisenberg model in Eq.~(\ref{201})] \cite{rashid1}.
Introducing easy-axis anisotropy permits to diminish the quantum fluctuations 
by increasing the amount of the Ising interaction.
We consider the Hamiltonian
\begin{eqnarray}
\label{401}
H
&=&
J\sum_{\langle nm\rangle} \left( s^x_{n} s^x_{m}+ s^y_{n} s^y_{m}+\Delta  s^z_{n} s^z_{m}\right)
+ \nonumber\\
&&
J^\prime\sum_{\langle nm\rangle^\prime} \left( s^x_{n} s^x_{m}+ s^y_{n} s^y_{m}+\Delta  s^z_{n} s^z_{m}\right)
\end{eqnarray}
with $\Delta\ge 1$.
At $\Delta=1$ the model (\ref{401}) coincides with Eq.~(\ref{201}), 
and in the limit $\Delta\to\infty$ it yields the Ising model, 
i.e., a model  without quantum fluctuations. 
Using the variational state given in Eq.~(\ref{301}) 
and performing corresponding calculations along the lines illustrated in Sec.~\ref{sec3} 
we arrive at
\begin{eqnarray}
\label{402}
\frac{E(t)}{N}
=
-\frac{J^\prime}{2}\frac{t}{1+t^2}
-\frac{\Delta J^\prime}{8}
-\frac{3\Delta J}{8} \left(\frac{1-t^2}{1+t^2}\right)^2,
\end{eqnarray}
cf. Eq.~(\ref{304}).
The variational energy (\ref{402}) is minimal for
\begin{eqnarray}
\label{403}
t
=
\left\{
\begin{array}{ll}
\frac{3\Delta J}{J^\prime} -\sqrt{\left(\frac{3\Delta J}{J^\prime}\right)^2-1}, & J^\prime\le 3\Delta J, \\
1,                                                          & J^\prime > 3\Delta J,
\end{array}
\right.
\end{eqnarray}
cf. Eq.~(\ref{308}),
and the ground-state energy (per site) is given by
\begin{eqnarray}
\label{404}
\frac{E_0}{N}
=
\left\{
\begin{array}{ll}
-\frac{(2+\Delta)J^\prime}{8}-\frac{3\Delta J}{8}\left(1 - \frac{J^\prime}{3\Delta J}\right)^2, & J^\prime\le 3\Delta J, \\
-\frac{(2+\Delta)J^\prime}{8},                                                                   & J^\prime > 3\Delta J,
\end{array}
\right.
\end{eqnarray}
cf. Eq.~(\ref{309}).
For the order parameter,
instead of Eq.~(\ref{312}), 
we have 
\begin{eqnarray}
\label{405}
m_0
=
\left\{
\begin{array}{ll}
\frac{1}{2} \sqrt{\left(1+\frac{J^\prime}{3J\Delta}\right)\left(1-\frac{J^\prime}{3J\Delta}\right)}, & J^\prime\le 3\Delta J, \\
0,                            & J^\prime > 3\Delta J.
\end{array}
\right.
\end{eqnarray}
Finally,
the Landau-like variational ground-state energy (per site) becomes
\begin{eqnarray}
\label{406}
\frac{E(m)}{N}
=
-\frac{2+\Delta}{8}J^\prime 
\nonumber\\
+ \frac{1}{2}\left(J^\prime -3\Delta J\right)m^2+\frac{1}{2}J^\prime m^4 +\ldots
\end{eqnarray}
cf. Eq.~(\ref{314}).

As can be seen from the reported formulas (\ref{404}), (\ref{405}), and (\ref{406}),
the quantum phase transition point $J_c^\prime$ is proportional to $\Delta$, 
i.e., $J_c^\prime/J$ tends to infinity in the Ising limit $\Delta\to\infty$.
With increase of the Ising anisotropy $\Delta$ the quantum fluctuations are reduced, 
thus pushing $J_c^\prime$  to higher values.
In the pure Ising limit
the only remaining in the Hamiltonian (\ref{401}) terms are the Ising interactions 
and the quantum critical point for transition into singlet-product state disappears: 
$J_c^\prime/J\to\infty$.
In other words,
in the pure Ising limit the ground state is of the N\'{e}el type for all $J^\prime/J$.
For further details see Ref.~\onlinecite{rashid1}.

\subsection{The $J-J^\prime$ model for higher spin quantum numbers $s>1/2$}

Another classical limit occurs when the spin value increases.
Let us discuss briefly how to treat higher spin quantum numbers $s=1,\,3/2,\,2,\ldots\,$.
For simplicity, we consider here the case $s=1$ only 
and refer the interested reader to Ref.~\onlinecite{rashid2}, 
where the cases $s=1$, $s=3/2$, and $s=2$ are discussed.
For $s=1$,
at each site one has three possible spin states:
$\vert1,1\rangle\equiv\vert\uparrow\rangle$,
$\vert1,0\rangle\equiv\vert\ 0 \rangle$,
and
$\vert1,-1\rangle\equiv\vert\downarrow\rangle$.
If $J^\prime=J$ we again have semiclassical two-sublattice N\'{e}el-type order 
where all spins, say, on the sublattice $A$ 
tend to be the spin-up state $\vert\uparrow\rangle$,
and all spins on the sublattice $B$ 
are preferably the spin-down state $\vert\downarrow\rangle$.
If $J^\prime\gg J$ we again expect the singlet state on the $J^\prime$ bonds.
However, now the singlet state is composed of two spins with $s=1$.
One can easily find the singlet eigenstate in the subspace with $S^z=0$,
i.e., among the eigenstates like 
$a\vert\uparrow_1\downarrow_2\rangle+b\vert 0_1 0_2\rangle+c\vert\downarrow_1\uparrow_2\rangle$,
where the coefficients $a$, $b$, and $c$ are to be found.
Using the relations
$s^+\vert\uparrow\rangle=0$,
$s^+\vert 0\rangle=\sqrt{2}\vert \uparrow\rangle$,
$s^+\vert \downarrow\rangle=\sqrt{2}\vert 0\rangle$,
$s^-\vert\uparrow\rangle=\sqrt{2}\vert 0\rangle$,
$s^-\vert 0\rangle=\sqrt{2}\vert \downarrow\rangle$,
and
$s^-\vert \downarrow\rangle=0$,
one finds the following three eigenstates of the operator ${\bf{s}}_1\cdot {\bf{s}}_2$:
$\vert\uparrow_1\downarrow_2\rangle-\vert 0_10_2 \rangle +\vert\downarrow_1\uparrow_2\rangle$ (singlet) with the energy $-2$,
$-\vert\uparrow_1\downarrow_2\rangle + 2\vert\downarrow_1\uparrow_2\rangle$ (triplet) with the energy $-1$,
and
$\vert\uparrow_1\downarrow_2\rangle+\vert 0_10_2 \rangle +\vert\downarrow_1\uparrow_2\rangle$ (quintuplet) with the energy $1$.

Bearing this in mind,
we modify accordingly the variational state in Eq.~(\ref{301})
and assume
\begin{eqnarray}
\label{407}
\vert\Psi_{\rm{var}}\rangle
=
\prod_{i\in A}\frac{1}{\sqrt{1+t_1^2+t_2^2}}
\nonumber\\
\times
\left(
\vert \uparrow_i \downarrow_{i+\hat{x}} \rangle 
- t_1 \vert 0_i 0_{i+\hat{x}} \rangle
+ t_2 \vert \downarrow_i \uparrow_{i+\hat{x}} \rangle
\right),
\end{eqnarray}
where $t_1$ and $t_2$ are the variational parameters.
For $t_1=t_2=0$ Eq.~(\ref{407}) reproduces the N\'{e}el state 
and 
for $t_1=t_2=1$ it gives the singlet-product state.
The variational ground-state energy (per site) and the variational sublattice magnetization (per site) are calculated as
\begin{eqnarray}
\label{408}
\frac{E(t_1,t_2)}{N}
=\frac{\langle \Psi_{\rm{var}} \vert H \vert \Psi_{\rm{var}} \rangle}{N}
\nonumber\\
=
-\frac{J^\prime}{2}\frac{1+2t_1+2t_1t_2+t_2^2}{1+t_1^2+t_2^2}
-\frac{3J}{2}
\left(\frac{1-t_2^2}{1+t_1^2+t_2^2}\right)^2
\end{eqnarray}
and
\begin{eqnarray}
\label{409}
m(t_1,t_2)=\langle \Psi_{\rm{var}} \vert s_1^z \vert \Psi_{\rm{var}} \rangle
=\frac{1-t_2^2}{1+t_1^2+t_2^2}.
\end{eqnarray}

Now we have to minimize $E(t_1,t_2)/N$ (\ref{408}) 
with respect to two variational parameters, $t_1$ and $t_2$.
We obtain a set of two coupled equations for $t_1$ and $t_2$ which can be solved numerically.
Numerics yield $J_c^\prime/J=8$ and $\beta=1/2$.
By inspecting  the critical value $J_c^\prime/J$ for $s=1/2,\,1,\,3/2,\,2$ in Ref.~\onlinecite{rashid2} the relation
\begin{eqnarray}
\label{410}
\frac{J^\prime_c}{J}=4s(s+1)
\end{eqnarray}
was found.
This scaling law for $J_c^\prime$
[as $s(s+1)$]
has been confirmed by other methods 
such as the coupled cluster method \cite{rashid2} or the bond-operator approach \cite{danu}.
As can be seen from Eq.~(\ref{410}),
in the classical limit $s\to\infty$ the quantum phase transition to the singlet-product state disappears:
$J_c^\prime/J\to\infty$ and the N\'{e}el state persists for any finite $J^\prime$.

\subsection{$J-J^\prime - J_2$ model}

\begin{figure}
\begin{center}
\includegraphics[clip=on,width=60mm,angle=0]{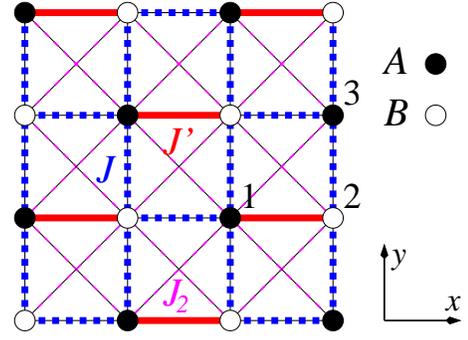}
\caption
{(Color online) 
Square-lattice $J-J^\prime - J_2$ model.}
\label{fig07}
\end{center}
\end{figure}

A further extension of the basic model (\ref{201}) 
is given by introducing a frustrating  antiferromagnetic next-nearest-neighbor interaction $J_2$.
Note that in the presence of frustration 
the powerful quantum Monte Carlo method used for the basic (unfrustrated) model (\ref{201}), 
see Refs.~\onlinecite{wenzel,fritz}, 
is not applicable because of the sign problem \cite{TrWi05}.
  
The Hamiltonian of the $J-J^\prime - J_2$ model reads
\begin{eqnarray}
\label{411}
H
=
J\sum_{\langle nm\rangle} {\bf{s}}_n\cdot {\bf{s}}_m 
+
J^\prime\sum_{\langle nm\rangle^\prime} {\bf{s}}_n\cdot {\bf{s}}_m 
\nonumber\\
+
J_2\sum_{\langle nm\rangle_2} {\bf{s}}_n\cdot {\bf{s}}_m, 
\end{eqnarray}
where the third sum in the r.h.s. of Eq.~(\ref{411}) runs over all next-nearest-neighbor bonds with the strength $J_2\ge 0$,
see Fig.~\ref{fig07}.
Now, in addition to calculations (\ref{302}) and (\ref{303}), 
we have to find
\begin{eqnarray}
\label{412}
&& \left(s_1^+s_3^- + s_1^+s_3^-\right)\vert \Psi_{\rm{var}} \rangle =
\nonumber\\
&&\quad
\ldots
\left(-t\right)\frac{1}{\sqrt{1+t^2}}\vert\uparrow_1\uparrow_2\rangle 
\frac{1}{\sqrt{1+t^2}}\vert\downarrow_3\downarrow_{3+\hat{x}}\rangle
\ldots +
\nonumber\\
&& \quad
\ldots
\left(-t\right)
\frac{1}{\sqrt{1+t^2}}\vert\downarrow_1\downarrow_2\rangle 
\frac{1}{\sqrt{1+t^2}}\vert\uparrow_3\uparrow_{3+\hat{x}}\rangle
\ldots,
\nonumber\\
&&s^z_1s^z_3\vert \Psi_{\rm{var}} \rangle =
\nonumber\\
&&\quad
\ldots
\left(+\frac{1}{4}\right)
\frac{1}{\sqrt{1+t^2}}\left(\vert\uparrow_1\downarrow_2\rangle + t\vert\downarrow_1\uparrow_2\rangle\right)
\times
\nonumber\\
&& \quad
\frac{1}{\sqrt{1+t^2}}\left(\vert\uparrow_3\downarrow_{3+\hat{x}}\rangle + t\vert\downarrow_3\uparrow_{3+\hat{x}}\rangle\right)
\ldots,
\nonumber\\
&& \langle \Psi_{\rm{var}} \vert {\bf{s}}_1\cdot{\bf{s}}_3 \vert \Psi_{\rm{var}} \rangle
=
+\frac{1}{4}\left(\frac{1-t^2}{1+t^2}\right)^2,
\end{eqnarray}
see Fig.~\ref{fig07}.
Importantly,
the sign of $\langle \Psi_{\rm{var}} \vert {\bf{s}}_1\cdot{\bf{s}}_3 \vert \Psi_{\rm{var}} \rangle$, see Eq.~(\ref{412}), 
is opposite to the sign of $\langle \Psi_{\rm{var}} \vert {\bf{s}}_2\cdot{\bf{s}}_3 \vert \Psi_{\rm{var}} \rangle$, see Eq.~(\ref{303}).
Summing the contributions of all bonds, 
i.e., of $N/2$ $J^\prime$ bonds, $3N/2$ $J$ bonds, and $2N$ $J_2$ bonds,
we arrive at
\begin{eqnarray}
\label{413}
\frac{E(t)}{N}=\frac{\langle \Psi_{\rm{var}} \vert H \vert \Psi_{\rm{var}} \rangle}{N}
\nonumber\\
=
-\frac{J^\prime}{2}\frac{t}{1+t^2}
-\frac{J^\prime}{8}
-\frac{3 J}{8} \left(\frac{1-t^2}{1+t^2}\right)^2
\nonumber\\
+\frac{ J_2}{2} \left(\frac{1-t^2}{1+t^2}\right)^2,
\end{eqnarray}
cf. Eq.~(\ref{304}).
Clearly, the last two terms in Eq.~(\ref{413}) can be combined
and after introducing the effective interaction
\begin{eqnarray}
\label{414}
J_{{\rm{eff}}}=J-\frac{4}{3}J_2
\end{eqnarray}
Eq.~(\ref{413}) becomes identical to Eq.~(\ref{304}) with $J_{{\rm{eff}}}$ (\ref{414}) instead of $J$.
For the critical point we get 
$J_c^\prime =3 J_{{\rm{eff}}}=3 J -4 J_2$.
Clearly,
the frustrating  coupling $J_2$ suppresses $J_c^\prime$ 
and acts in favor of the magnetically disordered singlet-product state.

Furthermore, 
for the case of $J_1-J_2$ model when $J^\prime=J=J_1$ 
the value of $J_2$ where the N\'{e}el order gives way for the valence-bond state
is $J^{c1}_2/J_1=1/2$.
This value is not far from $J^{c1}_2/J_1=0.4 \ldots 0.45$ 
obtained by more sophisticated methods for the critical frustration 
where the N\'{e}el order breaks down in the $J_1-J_2$ model \cite{darradi2008,j1j2,gong2014}.
Moreover,  
recent calculations using density matrix renormalization group approach with explicit implementation of $SU(2)$ spin rotation symmetry in Ref.~\onlinecite{gong2014} 
have found a gapless spin-liquid state for $0.44 < J_2/J_1 < 0.5$ 
and the transition to a gapped valence-bond phase takes place only at $J_2/J_1 =0.5$.  

\begin{figure}
\begin{center}
\includegraphics[clip=on,width=80mm,angle=0]{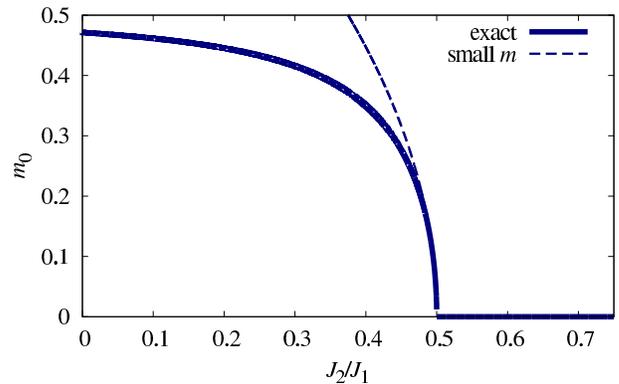}
\caption
{(Color online) 
The order parameter $m_0$ for the $J_1-J_2$ model 
(i.e., at $J=J^\prime=J_1$) 
calculated within the mean-field approach (\ref{301}).
The N\'{e}el order breaks down at $J^{c1}_2/J_1 = 1/2$.}
\label{fig08}
\end{center}
\end{figure}

The order parameter $m_0$ (sublattice magnetization per site) as a function of $J_2$ 
can be easily calculated by substituting $J^\prime/J\to J_1/(J_1-4J_2/3)$ in Eq.~(\ref{312}).
We show this dependence of $m_0$ on $J_2$ in Fig.~\ref{fig08}.

\subsection{The $J-J^\prime$ model on other lattices}

\begin{figure}
\begin{center}
\includegraphics[clip=on,width=60mm,angle=0]{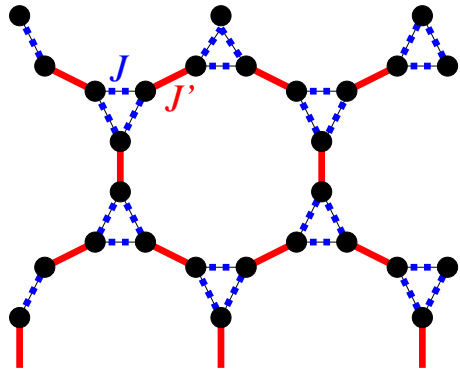}
\caption
{(Color online) 
Star-lattice model.
The star lattice has two different bonds:
dimer bonds $J^\prime$ and triangular bonds $J$.
For the uniform lattice these bonds are equal $J^\prime=J$.}
\label{fig09}
\end{center}
\end{figure}

The elaborated approach to examine quantum phase transitions in quantum Heisenberg antiferromagnets with competing bonds
can be straightforwardly applied to other lattices.
In case of unfrustrated lattices, 
e.g., the bilayer square-lattice \cite{gros}, 
for the isotropic $s=1/2$ model the calculations presented in Sec.~\ref{sec3} are still valid.
However, the number of nearest neighbors $z$ is a relevant parameter, 
cf. Eq.~(\ref{304}),
and $z$ has to be taken for the lattice under consideration. 
Thus, for the critical value we then have $J^\prime_c=(z-1)J$.
Another example  is the so-called CaVO (or 1/5-depleted square) lattice 
which is used to describe the magnetic properties of CaV$_4$O$_9$ \cite{lnp,arch_ccm,arch_Ising}.
For this lattice $z=3$ and therefore $J^\prime_c=2J$
that is in a reasonable agreement with quantum Monte Carlo data $J_c^\prime/J\approx 1.65$ \cite{troyer1996,capponi2009}.

\begin{figure}
\begin{center}
\includegraphics[clip=on,width=60mm,angle=0]{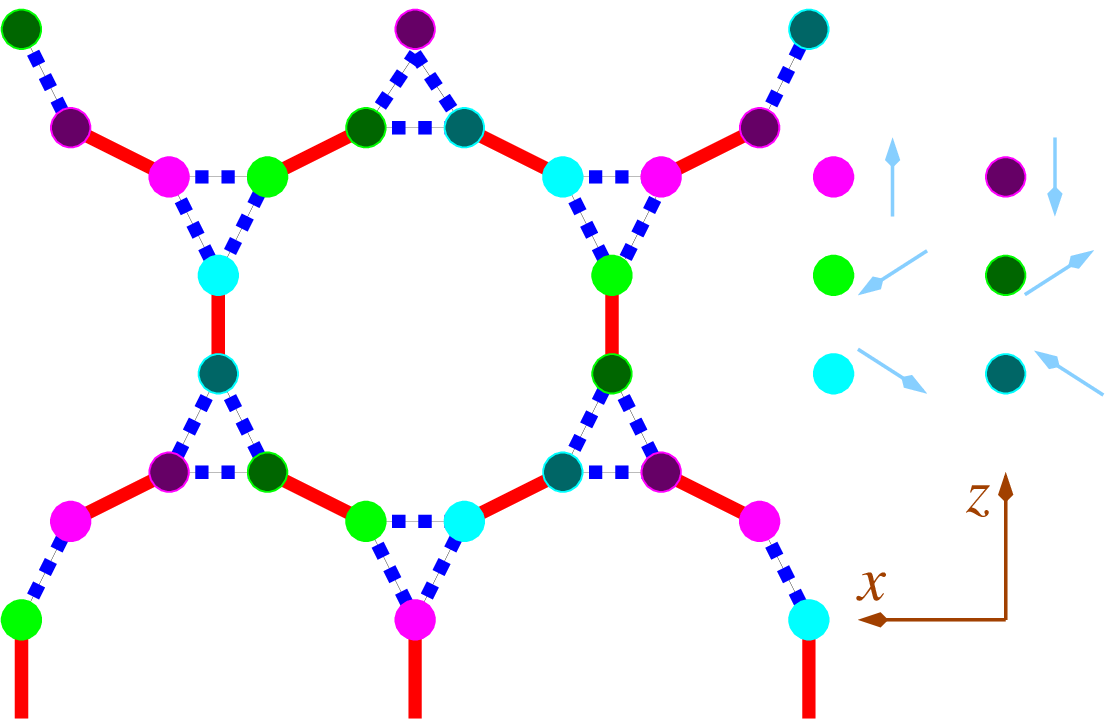}\\
\vspace{3mm}
\includegraphics[clip=on,width=60mm,angle=0]{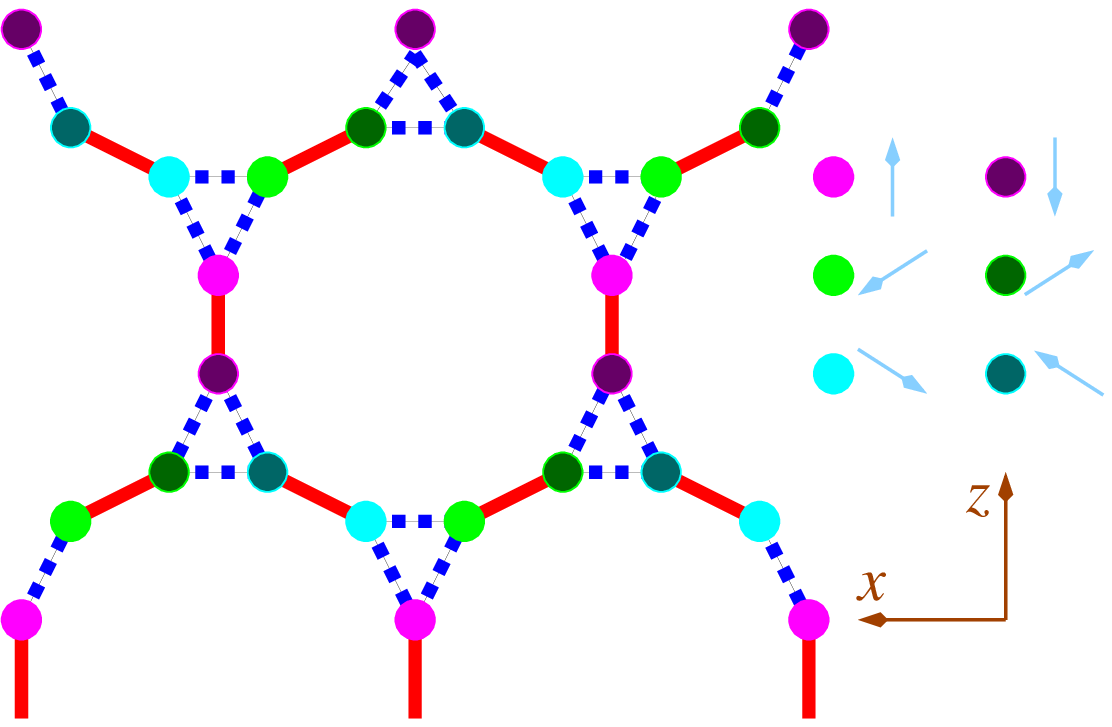}\\
\vspace{5mm}
\includegraphics[clip=on,width=60mm,angle=0]{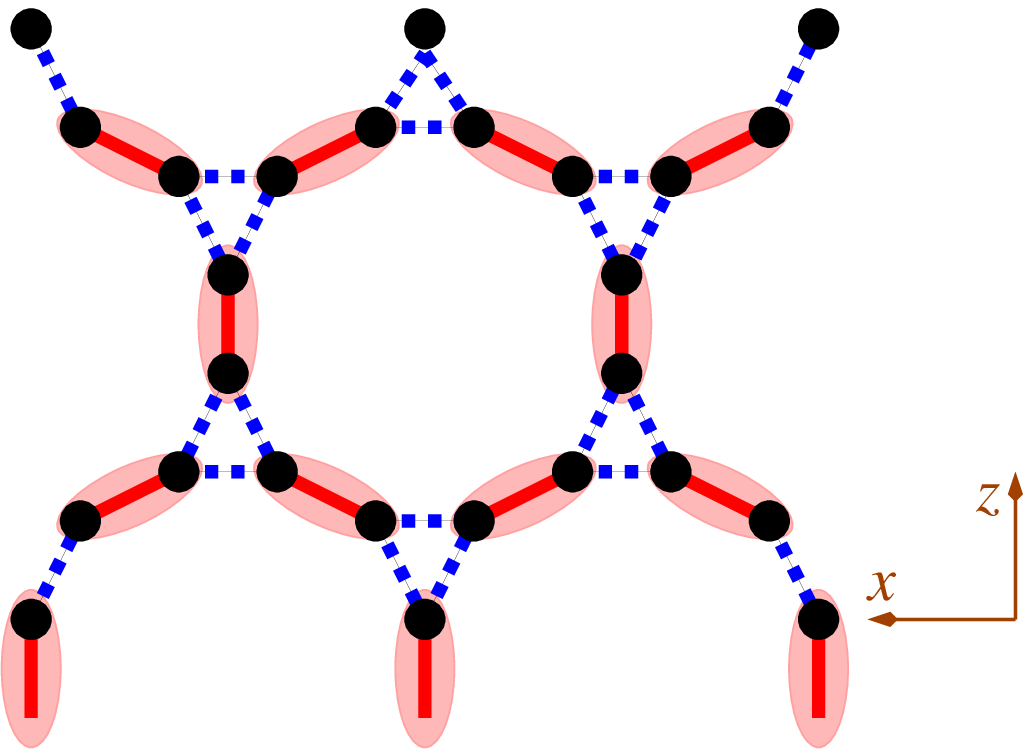}
\caption
{(Color online) 
Relevant states of the star-lattice model.
From top to bottom:
the $\sqrt{3}\times\sqrt{3}$ state,
the ${\bf{q}}=0$ state,
and
the singlet-product state.}
\label{fig10}
\end{center}
\end{figure}

A more interesting situation can appear on non-bipartite lattices, 
where due to geometrical frustrations the semiclassical magnetic order  typically is non-collinear.             
As an example, we consider the star lattice, 
see Fig.~\ref{fig09}.
This more exotic lattice is one of the 11 uniform Archimedean tilings in dimension $D=2$ \cite{lnp,arch_ccm,arch_Ising}. 
The Hamiltonian of the model is given in Eq.~(\ref{201}),
however, now the first sum runs over all triangular bonds and the second one over all dimer bonds,
see Fig.~\ref{fig09}.
For the uniform lattice both bonds, the triangular and dimer ones, have the same strength.
In this case, two variants of a semiclassical ground state were discussed 
(analogs of the N\'{e}el state for the square lattice):
i) the so-called $\sqrt{3}\times\sqrt{3}$ state
and
ii) the so-called ${\bf{q}}=0$ state,
see the two upper panels in Fig.~\ref{fig10}.
These states should appear in Eq.~(\ref{301}) instead of the N\'{e}el state.

The calculations for the star-lattice model can be sketched as follows.
We consider either the $\sqrt{3}\times\sqrt{3}$ state or the ${\bf{q}}=0$ state,
see Fig.~\ref{fig10}.
Since ${\bf{s}}_n\cdot{\bf{s}}_m$ is rotationally invariant in the spin space,
we may choose the $z$ axis in the spin space within the plane defined by the $\sqrt{3}\times\sqrt{3}$ or ${\bf{q}}=0$ state 
(both are coplanar states).
Then, we rotate the spins around the $y$ axis by an angle $\alpha$ which is site dependent, 
i.e., we perform local rotations
\begin{eqnarray}
\label{415}
s^x\to \tilde{s}^x= \cos\alpha\, s^x -\sin\alpha\, s^z ,
\nonumber\\
s^y\to \tilde{s}^y=s^y,
\nonumber\\
s^z\to \tilde{s}^z= \sin\alpha\, s^x + \cos\alpha\, s^z
\end{eqnarray}
with
$\alpha=\pi$ for ``magenta'' sites,
$\alpha=-\pi/3$ for ``green'' sites,
and
$\alpha=\pi/3$ for ``cyan'' sites,
see Fig.~\ref{fig10}.
Clearly,
the spin state at all sites shown in the two upper panels of Fig.~\ref{fig10}
is now the eigenstate of the operator $\tilde{s}^z$ 
(in the rotated coordinate frame) 
with the eigenvalues $+1/2$ (dark sites) or $-1/2$ (light sites).
We denote them $\vert\tilde{\uparrow}\rangle$ and $\vert\tilde{\downarrow}\rangle$, respectively.
Note, however,
that in the rotated coordinates the Hamiltonian becomes more complicated, 
since
\begin{eqnarray}
\label{416}
{\bf{s}}_n\cdot{\bf{s}}_m =
\nonumber\\
\cos\left(\alpha_n-\alpha_m\right) \left(\tilde{s}_n^x \tilde{s}_m^x +  \tilde{s}_n^z \tilde{s}_m^z\right)
+\tilde{s}_n^y \tilde{s}_m^y
\nonumber\\
-\sin\left(\alpha_n-\alpha_m\right) \left(\tilde{s}_n^x\tilde{s}_m^z-\tilde{s}_n^z\tilde{s}_m^x\right).
\end{eqnarray}

Now we adopt a variational state rewriting Eq.~(\ref{301}) in the form:
\begin{eqnarray}
\label{417}
\vert\Psi_{\rm{var}}\rangle
=
\prod_{\langle nm\rangle^\prime}\frac{1}{\sqrt{1+t^2}}
\left(
\vert \tilde{\uparrow}_n \tilde{\downarrow}_{m} \rangle - t \vert \tilde{\downarrow}_n \tilde{\uparrow}_{m} \rangle
\right),
\end{eqnarray}
where the product runs over all dimer bonds $J^\prime$ of the star lattice.
Next, we have to calculate the contribution to the variational energy from different bonds.
In total there are $3N/2$ bonds: $N/2$ of them are dimer bonds and $N$ of them are triangular bonds.
The contribution of the dimer bonds is given by Eq.~(\ref{302}).
The contribution of the triangular bond is given by Eq.~(\ref{303}) 
with taking into account the factor 1/2 stemming from $\cos(\alpha_n-\alpha_m)=1/2$ in Eq.~(\ref{416}).
As a result,
we obtain for the variational energy $E(t)$ the following formula:
\begin{eqnarray}
\label{418} 
\frac{E(t)}{N}
=
-\frac{J^\prime}{2} \left(\frac{t}{1+t^2}+\frac{1}{4}\right)
-\frac{J}{8}\left(\frac{1-t^2}{1+t^2}\right)^2,
\end{eqnarray}
cf. Eq.~(\ref{304}).
The obtained variational energy implies $J_c=J$: 
One arrives at this outcome simply by comparing Eq.~(\ref{418}) and Eq.~(\ref{304}).
Thus for the uniform lattice the magnetic order is already unstable,
and, therefore, this result may serve as an indication of the absence of magnetic order for the star-lattice spin-1/2 Heisenberg antiferromagnet.
This result found by using the simple mean-field like variational approach 
is indeed in agreement with findings using more advanced many-body methods \cite{johannes,lnp,kim2010}.

\section{Conclusions and outlook}
\label{sec5}
\setcounter{equation}{0}

To summarize,
we have considered a mean-field like approach for the analysis of quantum phase transitions 
in quantum spin systems with competing antiferromagnetic bonds.
This scheme can be cast into the standard Landau's paradigm of phase transitions.
Furthermore,
the method is rather transparent and simple from the calculation point of view.
This method provides reasonably good estimates for quantum critical points,
and the critical behavior falls into the mean-field universality class.
Because of the local character of the variational ansatz (\ref{301}),
this method cannot distinguish between various distinct patterns for the arrangement of dimers 
and therefore it cannot provide more refined information on the features of the quantum phase transition.

\section*{Acknowledgments}

The authors are grateful to T.~Verkholyak and P.~M\"{u}ller for critical reading of the manuscript.
The present study was supported by the Deutsche Forschungsgemeinschaft (project RI615/21-2).
The work of O.~D. was partially supported by Project FF-30F (No.~0116U001539) from the Ministry of Education and Science of Ukraine.
O.~D. acknowledges the kind hospitality of the University of Magdeburg in March-May and October-December of 2016.
O.~D. acknowledges the kind hospitality of the University of Ko\v{s}ice during the CSMAG16 conference in June of 2016.
O.~D. would also like to thank the Abdus Salam International Centre for Theoretical Physics (Trieste, Italy)
for support of this study through the Senior Associate award in August of 2016.


\begin{thebibliography}{99}

\bibitem{t_sp1}
L.~D.~Landau and E.~M.~Lifshitz,
{\it Statistical Physics},
third revised and enlarged edition
(Pergamon Press, Oxford, 1980).

\bibitem{t_sp2}
R.~Balescu,
{\it Equilibrium and Nonequilibrium Statistical Mechanics}
(A Wiley-Interscience Publication, New York, 1975).

\bibitem{t_sp3}
L.~E.~Reichl,
{\it A Modern Course in Statistical Physics},
4th revised and updated edition
(Wiley-VCH Verlag GmbH\&Co.KGaA, 2016).

\bibitem{onsager}
L.~Onsager,
Phys. Rev. {\bf 65}, 117 (1944);
T.~D.~Schultz, D.~C.~Mattis, and E.~H.~Lieb,
Rev. Mod. Phys. {\bf 36}, 856 (1964).

\bibitem{sachdev}
S.~Sachdev,
Science {\bf 288}, 475 (2000);
S.~Sachdev,
{\it Quantum Phase Transitions},
second edition
(Cambridge University Press, 2011). 

\bibitem{voita}
M.~Vojta,
Reports on Progress in Physics {\bf 66}, 2069 (2003).

\bibitem{derzhko}
O.~Derzhko,
in
{\it Order, Disorder and Criticality: Advanced Problems of Phase Transition Theory},
edited by Yurij Holovatch
(World Scientific, Singapore, 2004),
pp.~109-145. 

\bibitem{bitko}
D.~Bitko, T.~F.~Rosenbaum, and G.~Aeppli,
Phys. Rev. Lett. {\bf 77}, 940 (1996).

\bibitem{buch1}
{\it Quantum Magnetism}, 
Lecture Notes in Physics {\bf{645}}, 
edited by U.~Schollw\"{o}ck, J.~Richter, D.~J.~J.~Farnell, and R.~F.~Bishop
(Springer-Verlag, Berlin, Heidelberg, 2004).

\bibitem{buch2}
{\it Introduction to Frustrated Magnetism}, 
Springer Series in Solid-State Sciences \textbf{164},
edited by C.~Lacroix, P.~Mendels, and F.~Mila,
(Springer-Verlag, Berlin, Heidelberg, 2011).

\bibitem{ritz}
W.~Ritz,
Journal f\"{u}r die reine und angewandte Mathematik {\bf 135}, 1 (1909).
   
\bibitem{c_gros}
C.~Gros,
Annals of Physics (N.Y.) {\bf 189}, 53 (1989).

\bibitem{becca}
F.~Becca, L.~Capriotti, A.~Parola, and S.~Sorella,
in 
{\it Introduction to Frustrated Magnetism},
Springer Series in Solid-State Sciences \textbf{164},
edited by C.~Lacroix, P.~Mendels, and F.~Mila,
(Springer-Verlag, Berlin, Heidelberg, 2011),
pp.~379-406.

\bibitem{wenzel}
S.~Wenzel, L.~Bogacz, and W.~Janke,
Phys. Rev. Lett. {\bf 101}, 127202 (2008);
S.~Wenzel and W.~Janke,
Phys. Rev. B {\bf 79}, 014410 (2009).

\bibitem{fritz}
L.~Fritz, R.~L.~Doretto, S.~Wessel, S.~Wenzel, S.~Burdin, and M.~Vojta,
Phys. Rev. B {\bf 83}, 174416 (2011).

\bibitem{manousakis}
E.~Manousakis,
Rev. Mod. Phys. {\bf 63}, 1 (1991).

\bibitem{oitmaa1992}
C.~J.~Hamer, Zheng Weihong, and P.~Arndt,
Phys. Rev. B {\bf{46}}, 6276 (1992).

\bibitem{darradi2008}
R.~Darradi, O.~Derzhko, R.~Zinke, J.~Schulenburg,  S.~E.~Kr\"{u}ger, and J.~Richter,
Phys. Rev. B {\bf 78}, 214415 (2008).

\bibitem{lnp}
J.~Richter, J.~Schulenburg, and A.~Honecker, 
in {\it Quantum Magnetism}, 
Lecture Notes in Physics {\bf{645}},
edited by U.~Schollw\"{o}ck, J.~Richter, D.~J.~J.~Farnell, and R.~F.~Bishop 
(Springer-Verlag, Berlin, Heidelberg, 2004), 
pp.~85-153.

\bibitem{gros}
C.~Gros, W.~Wenzel, and J.~Richter,
Europhys. Lett. {\bf 32}, 747 (1995). 

\bibitem{sven}
S.~E.~Kr\"{u}ger, J.~Richter, J.~Schulenburg, D.~J.~J.~Farnell, and R.~F.~Bishop,
Phys. Rev. B {\bf 61}, 14607 (2000).

\bibitem{johannes}
J. Richter, J. Schulenburg, A. Honecker, and D. Schmalfu\ss ,
Phys. Rev. B {\bf 70}, 174454 (2004).

\bibitem{rashid1}
R.~Darradi, J.~Richter, and S.~E.~Kr\"{u}ger,
Journal of Physics: Condensed Matter {\bf 16}, 2681 (2004).

\bibitem{rashid2}
R.~Darradi, J.~Richter, and D.~J.~J.~Farnell,
Journal of Physics: Condensed Matter {\bf 17}, 341 (2005).

\bibitem{landau1937}
L.~Landau,
Zh. Eksp. Teor. Fiz. {\bf 7}, 19 (1937)
[see also
Ukr. J. Phys. {\bf 53}, Special Issue, 25 (2008)].

\bibitem{danu}
B.~Danu and B.~Kumar,
arXiv:1610.03826v1.

\bibitem{TrWi05} 
M.~Troyer and U.-J.~Wiese,
Phys. Rev. Lett. {\bf 94}, 170201 (2005).

\bibitem{j1j2}
J.~Sirker, Zheng~Weihong, O.~P.~Sushkov, and J.~Oitmaa,
Phys. Rev. B {\bf 73}, 184420 (2006).

\bibitem{gong2014} 
Shou-Shu~Gong, Wei~Zhu, D.~N.~Sheng, O.~I.~Motrunich, and M.~P.~A.~Fisher, 
Phys. Rev. Lett. {\bf 113}, 027201 (2014).

\bibitem{arch_ccm}  
D.~J.~J.~Farnell, O.~G\"{o}tze,  J.~Richter, R.~F.~Bishop, and P.~H.~Y.~Li,
Phys. Rev. B {\bf 89}, 184407 (2014).

\bibitem{arch_Ising} 
Unjong~Yu,
Phys. Rev. E {\bf 91}, 062121 (2015).

\bibitem{troyer1996}
M.~Troyer, H.~Kontani, and K.~Ueda,
Phys. Rev. Lett. {\bf 76}, 3822 (1996).

\bibitem{capponi2009}
D.~Schwandt, F.~Alet, and S.~Capponi,
Phys. Rev. Lett {\bf 103}, 170501 (2009).

\bibitem{kim2010}
B.-J.~Yang, A.~Paramekanti, and Y.~B.~Kim,
Phys. Rev. B {\bf 81}, 134418 (2010).  

\end{thebibliography}
\end{document}